\documentclass[prb,onecolumn,12pt]{revtex4-1}
\usepackage{amssymb}
\usepackage{amsmath}
\usepackage{graphicx}
\usepackage{color}
\usepackage[normalem]{ulem}
\normalsize

\usepackage[latin2]{inputenc}

\newcommand{\pder}[3]{\frac{\partial^{#1}{}#2}{\partial{}#3^{#1}}} 

\begin{document}


\title{A non-equilibrium thermodynamics model for combined adsorption and diffusion processes
in micro- and nanopores}

\author{I. Santamar\'{\i}a-Holek$^1$\footnote{Permanent address: UMJ-Facultad de Ciencias, 
Universidad Nacional Aut\'onoma de M\'exico Campus Juriquilla, Bvd. Juriquilla 3001, C.P. 76230, 
Quer\'etaro, Mexico}, Z.J. Grzywna,$^2$ J. M. Rubi,$^1$}
\affiliation{ 
$^1$ Facultat de F\'isica, Universitat de Barcelona, Av. Diagonal 647, 08028 Barcelona, Spain}
\affiliation{$^2$ Department of Physical Chemistry and Technology of Polymers, Silesian University of Technology, Strzody 9, 44-100 Gliwice, Poland}

\begin{abstract}
A non-equilibrium thermodynamics model able to analyze the combined effect of diffusion and adsorption 
in porous materials is proposed. The model considers the coupled dynamics of the diffusive phase, 
described by a diffusion type equation, and the adsorbed phase governed by a generalized Langmuir 
equation. It is shown that the combination of diffusion and adsorption can be treated as diffusion 
in an effective medium, with an effective diffusion coefficient depending on the characteristics 
of the adsorption 
kinetics. The interplay between these effects leads to the appearance of different regimes in which the 
effective diffusion coefficient exhibits peculiar behaviours such as a decrease due to the 
presence of the adsorption process and a non-monotonous behavior resulting from particle 
interactions. We outline applications of the model developed to diffusion in porous materials such 
as zeolites and bentonites, and to engineered highly porous materials.
\end{abstract}
\maketitle

\section*{1. Introduction}
Diffusion is a fundamental irreversible process intervening in the evolution of many out of equilibrium 
systems. At sufficiently large scales it is successfully described by Fick's 
law obtained from non-equilibrium thermodynamics \cite{mazur} and, at shorter time scales,
by means of kinetic equations of the Fokker-Planck type derived from mesoscopic non-equilibrium 
thermodynamics \cite{review,inertial,rubi01}.  

When the particles diffuse through a heterogeneous or complex medium, the structure and
the physicochemical properties of the medium may give rise to the appearance of additional kinetic 
processes that affect the dynamics of the particles\cite{paul,smit,hu,saxena,wloch,papa}. An important example of this situation, 
with applications ranging from biomedicine to engineering,
is diffusion in the presence of adsorption in which particles moving through the pores of the material have certain 
probability to become attached at the walls of the pores \cite{baranowski,rutven,roque,nanochemistry,karger,brinkmann,fillipova,gaspard,vazquez,grzywna,brandani,martin}. 
Two possibilities can be analyzed: total or partial immobilization of
the particles at the surface \cite{brinkmann,jost,vazquez,grzywna}. The kinetics is 
thus dominated by a set of transport equations that may present 
different relaxation time scales depending on the physicochemical 
properties of the surface and of the bulk fluid \cite{granick-science,brinkmann,gaspard}.

Our purpose in this article is to propose a simple model able to analyze the mutual influence of diffusion and adsorption processes 
in micro and nanopores, a situation observed in zeolites, bentonites or in new engineered highly porous materials 
\cite{simon,jost,vazquez,brandani,bentonita, natureRev}. We are interested in provide a simple theoretical model able
to explain, at least in qualitative form, how this mutual influence is responsible for interesting
effects observed in experiments \cite{rutven} and recently studied via molecular dynamics and cellular automaton techniques 
on the diffusion properties of the particles in the pore \cite{suffriti1,suffriti2,suffriti3}.
Examples of these effects are the non-monotonic behavior of the diffusivity as a function of particle concentration or particle loading of the pore, 
or even the strong suppression of self-diffusivity as the number of particles in the pore increases \cite{suffriti1,suffriti2,suffriti3}.
The model describes diffusion of one component in another at rest by means of the Fick law and adsorption through 
a generalized Langmuir equation obtained from non-equilibrium thermodynamics \cite{rubi01}. 
Using the fact that the concentration of the diffusing particles depends on that of the adsorbed 
particles, due to mass conservation, we formulate an effective medium theory in which 
the overall process is described as 
effective diffusion, with a diffusion coefficient depending on the characteristics of the 
adsorption process. Our analysis on the effects of adsorption on bulk diffusion is complementary to the usual theoretical 
approach to the problem which makes emphasis on the surface dynamics of the adsorbed 
phase \cite{baranowski,rutven,roque,brinkmann,fillipova,gaspard}.Here,
we are considering the case of isothermal adsorption. The generalization to the case when non-isothermal effects, such as pervaporation 
associated to the mass transport through a zeolite membrane could also be performed by following a non-equilibrium thermodynamics
analysis similar to that of Ref. [30]. 

The effective diffusion coefficient is calculated for a particular adsorption
kinetics. It is shown that it includes a factor accounting for its 
deviation for the case of absence of adsorption. At high concentrations of 
adsorbed particles, a virial expansion is proposed to estimate the effects of 
interactions. We conclude that the effective transport properties of the system
strongly depend on the adsorption-desorption rate constants as well as on temperature through the
corresponding virial coefficients. The great multiplicity of behaviours found 
reveals the strong influence that one of these proceses exerts on the other. 

The paper is organized as follows. In Section 2, we present the model and analyze the case 
of total immobilization of the adsorbed phase. Section 3 is devoted to 
study the effect of surface diffusion of the adsorbed phase on the diffusion/adsorption process. 
Finally, our conclusions are presented in Section 4.

\section*{2. Interplay between adsorption and diffusion}

Non-equilibrium thermodynamics provides a general framework to describe both 
adsorption and diffusion process and to analyze their mutual influence \cite{mazur,rubi01}.

In many instances, diffusion and adsorption occur simultaneously.
Particles in the bulk phase diffuse and arrive at the 
surface where they become attached due to 
the presence of molecular forces. This last process 
corresponds to adsorption and has been modeled by considering it as a diffusion 
through a potential barrier \cite{rubi01}. 

\begin{figure}[tbp] 
\begin{center}
\includegraphics[scale=0.40]{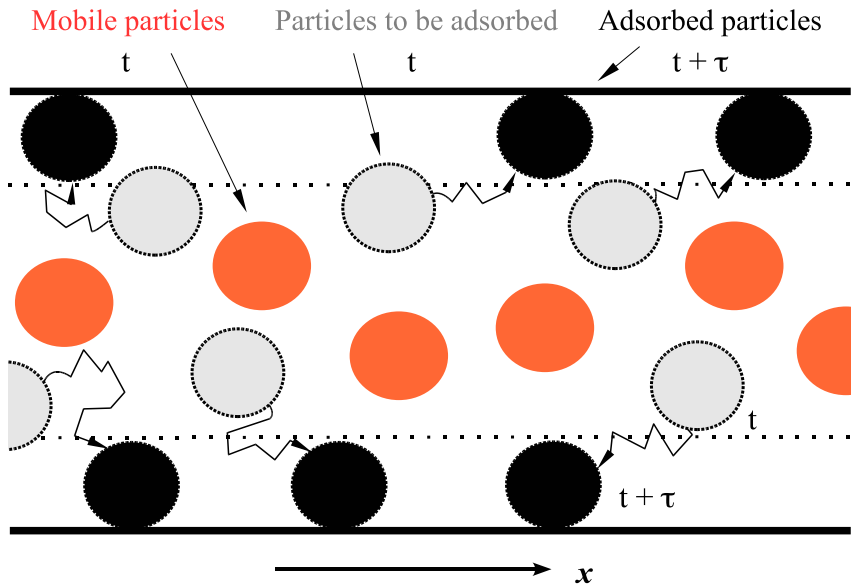} 
\caption{Diffusing particles in a quasi-onedimensional micropore. The orange circles represent 
the mobile particles in the porous at time $t$. The light grey circles represent the mobile particles at time $t$ 
that become adsorbed at time $t+\tau$. The lines schematically represent the trajectories they follow. 
The black circles represent the particles adsorbed at time $t+\tau$}.
\label{fig0}
\end{center}
\end{figure}

When the pore is very narrow, the particles practically move along the pore,
in a thin central layer  (see Fig. 1).  Under these conditions transport is 
quasi-onedimensional and the corresponding descripton can be performed in terms 
of averaged concentrations depending on the coordinate along the axis of the pore.
It is convenient to stress here that we will only consider the case when 
the diffusing particles have a coherent structure, such that their geometry can be characterized by an effective radius 
which is much larger than the molecules of the solvent. This assumption avoids to consider the rotational dynamics of the
particles with more complicated geometries.

After averaging the three dimensional diffusion equations over $(y,z)$-directions, 
it can be shown that the evolution equation for the average number concentration $c_1(x)$ in 
the bulk is given by \cite{mazur,rubi01}
\begin{equation}\label{consmasa1}
\frac{\partial c_1(x,t)}{\partial t}=-\frac{\partial}{\partial x} {J}_1(x,t)+j, 
\end{equation}
where $J_1$ is the diffusion current of particles having dimensions of 
number of particles per unit area and time along the $x$-direction,
and $j$ has dimensions of number of particles divided by volume and time and represents 
the net amount of particles leaving the bulk by crossing the dotted lines indicated in 
Figure 1.
In similar way,
the evolution equation for the concentration $c_2(x)$ of the adsorbed phase is
\begin{equation}\label{consmasa2}
\frac{\partial c_2(x)}{\partial t}=-\frac{\partial}{\partial x} J_2(x)-j+j_{cs}, 
\end{equation}
Here $J_2(x)$ is the corresponding diffusion current density and
we used the fact that the net amount of particles arriving at the surface is $j_{2n}=-j$
and that $j_{cs}$ is the reaction term associated to chemisorption. In the following, for
simplicity we will only consider the case of physical adsorption, that is, when  $j_{cs}=0$.
Further details on the derivation of Eq.  (\ref{consmasa2}) are given in Ref. [4].

The expression for the currents entering Eqs. (\ref{consmasa1}) and (\ref{consmasa2})
follows from the entropy production rate per unit mass $\sigma$  of the 
system \cite{rubi01}. 
This quantity is given by $\sigma=\sigma_1+\sigma_2$, where 
the contribution associated to the bulk is
\begin{equation}\label{sigmab}
\sigma_1 =  -\frac{1}{T} {J}_1 \frac{\partial}{\partial x} \mu_1 ,
\end{equation}
and the entropy production at the surface is given by \cite{rubi01}
\begin{equation}\label{sigmas}
\sigma_2 =  -\frac{1}{T} {J}_2 \frac{\partial}{\partial x} \mu_2
-\frac{1}{T}\,j\left(\mu_1-\mu_2\right),
\end{equation}
where $\mu_i(x,t)$ ($i=1,2$) represent the corresponding non-equilibrium chemical potentials. 
It is convenient to mention that, by hypothesis, no chemical reactions occur 
in the bulk and therefore Eq. (\ref{sigmab}) has no contribution due to chemical reactions.
From Eqs. (\ref{sigmab}) and (\ref{sigmas}) and assuming linear relationships between
fluxes and forces, one finds: 
$J_1 \propto \left({\partial}\mu_1/{\partial x}\right) $, 
$J_2 \propto \left({\partial}\mu_2/{\partial x}\right) $, 
$ j \propto \mu_1-\mu_2$. 

 In the dilute case at constant temperature, the last relation implies that the flow of particles 
from the bulk to the surface $j$ becomes proportional to the pressure difference which in turn is proportional to the 
concentration (assuming Henry's law in the bulk). This flow has two contributions, $j=j_a+j_d$, one for the particles leaving the bulk, $j_a$,
and one for the particles arriving to the bulk from the surface, $j_d$. The first contribution is proportional to a constant rate $\tilde{k}_a$ 
giving the probability per unit time that the particles in the bulk is adsorbed by the surface. However, this
probability must depend on the available surface. In Langmuir approach, this factor was assumed as
proportional to the concentration of particles at the surface: $1-c_2/c_{2}^{max} $ with $c_{2}^{max}$ the maximum of
adsorbing sites at the surface. However, if interactions
exist among these particles, then it depends upon the non-linear function $Z_2(c_2)$ which takes into
account excluded volume interactions of the adsorbed phase at the surface \cite{rubi01} .
That is, $\tilde{k}_a=k_a Z_2(c_2)$ with $Z_2=1-c_2/c_{2}^{max} $ for ideal systems and 
$Z_2(c_2)$ arbitrary in the general case. Hence, the flow $j_a$ can be written as:
$i$) $ j_{a}= - k_a c_1 (1-{c_2}/{ c_{2}^{max} })$ for Langmuir and 
$ii$) $j_{a}= - k_a c_1 Z_2(c_2)$ for the non-ideal case. In both expressions $k_a$ represents the adsorption rate constant.
The second contribution $j_d$ is proportional to the covered surface and is usually expressed as $j_d = - k_d c_2$, where
$k_d$ is the desorption rate constant. Using these relations in Eqs. (\ref{consmasa1}) and (\ref{consmasa2}) one obtains
\begin{equation}\label{R-D1}
\frac{\partial}{\partial t} c_1=\frac{\partial}{\partial x} \left[ D_1 \frac{\partial c_1}{\partial x} \right] +j,
\end{equation}
for the mass balance in the bulk, whereas for the adsorbed particles in the nonlinear case we have
\begin{equation}\label{R-D2}
\frac{\partial}{\partial t} c_2 = k_a c_1 Z(c_2) - k_dc_2  
+\frac{\partial}{\partial x}\left[ D_2 \frac{\partial c_2}{\partial x} \right].
\end{equation}
In writing Eqs. (5) and (6) we are considering the concentrations $c_1$ and $c_2$ in terms of volume. 
This allows us to use the mass conservation condition $c=c_1+c_2$, in order to considerably simplify the description with out 
loss of generality, and takes into account the fact that surface diffusion is strictly a three dimensional process that occurs in a 
small volume surrounding the surface of the solid. Thus, Eqs. (\ref{R-D1}) and (\ref{R-D2}), together with mass conservation 
$c=c_1+c_2$, describe the dynamics of the particles in the pore.  
It is convenient to mention that the adsorption rate constant follows an Arrhenius law in which this quantity 
depends on temperature Ref. 4. 
As a consequence, the interplay between adsorption and diffusion could be controlled by changing the temperature of the bath.

Instead of proceeding with the numerical solution of Eqs. (\ref{R-D1}) and (\ref{R-D2}), 
we will use mass conservation in order to adopt a simpler effective 
medium description for the interplay between adsorption and diffusion processes in which
the boundary term $j$ in Eq. (\ref{R-D1}) does not appears explicitly. 
This can be done because this term is coupled to the reaction term in (\ref{R-D2}), which in fact
determines a relation between the concentrations $c_1$ and $c_2$ corresponding to the adsorption isotherm.
For instance, in the stationary case with no surface diffusion effects, $j$ is strictly zero. However, even in this case,
we now that the diffusion of particles in the bulk is modified by the adsorption process, in such a way that 
$D_1=D_1(c_1,c_2)$. Using the mass conservation relation and the corresponding adsorption isotherm, this
effect can be incorporated through an effective diffusion coefficient $D_{eff}(c)$. \cite{mazur,jost,vazquez} 
The effective diffusion current  for the total number concentration is
\begin{equation} \label{immfluxtotal}
J_{eff}=-D_{eff}\frac{\partial c}{\partial x},
\end{equation}
Note that the effective diffusion coefficient $D_{eff}(c)$ can be associated to an effective activity 
$a_{eff}(c)=cf_{eff}(c)$ in such a way that we may write
\begin{equation} \label{termodDf}
{D}_{eff}(c)=RTB\left[1+\pder{}{ln|f_{eff}|}{ln|c|}\right], 
\end{equation}
where $f_{eff}(c)$ is the effective activity coefficient and $B$ is the mobility which in the present case can be
well approximated by the Stokes friction coefficient, $B=6\pi \eta R_{eff}$ where $\eta$ is the dynamic viscosity of the solvent
and $R_{eff}$ is the effective radius of the particle. Using equations (\ref{immfluxtotal}) 
and (\ref{termodDf}) in the mass conservation law for the bulk yields
\begin{eqnarray}\label{ogolne}
 \pder{}{c}{t}=\frac{\partial}{\partial x}\left[D_{eff}\frac{\partial c}{\partial x} \right]. 
\end{eqnarray}
It remains now to determine the expression of $D_{eff}(c)$ which, 
after using the adsorption kinetics and the mass conservation condition, can be expressed as a function of
the total concentration $c$. This will be 
performed by analyzing different cases in the next section.

\subsection*{2.1. Total immobilization  of the adsorbed phase}
The case in which adsorbed particles are completely immobilized
follows from the general model by imposing that the effective diffusion current $J_{eff}$, 
referring initially only to the mobile fraction $c_1$ is given by
\begin{equation}\label{immflux}
J_{eff}=-D_1(c_1,c_2)\pder{}{c_1}{x}.
\end{equation} 
The diffusion process can also be described by means of the effective Fick's law (\ref{immfluxtotal}).
This fact implies a relationship between the diffusion coefficients $D_1(c_1)$ and $D_{eff}(c)$
that can be established by considering the constraint $c=c_1+c_2$, with 
the subindex $2$ corresponding to the immobile fraction. Since $c_2$ depends on $c_1$ 
through the adsorption isotherm: $c_2=\phi(c_1)$, this gives 
$c=c_1+\phi(c_1)$ or equivalently $c_1=c_1(c)$. Thus, using the equality 
$\partial{c_1}/\partial{x}=\left(\partial{c_1}/\partial{c}\right)\partial{c}/\partial{x}$,
we finally obtain
\begin{equation}\label{imm_flux2}
D_{eff}(c)=\Delta(c) {D}_1(c),
\end{equation}
where the factor $\Delta(c)\equiv \partial{c_1}/\partial{c}$ accounts for deviations of the diffusion
coefficient with respect to its value in the case of absence of adsorption.

The influence of adsorption on diffusion enters through the evolution 
in time of the concentration of the immobilized particles $c_2$. An interesting analysis 
on how distinct kinetic regimes may appear in adsorption-diffusion interplay is discussed in
Ref. \cite{granick-science}. This dependence 
can be determined by means of  the generalized Langmuir adsorption equation \cite{rubi01}
\begin{equation}\label{kinetic-langmuir}
\frac{\partial}{\partial t} c_2 = k_a c_1 Z(c_2) - k_d c_2.
\end{equation}
When adsorption occurs in time scales much shorted than the characteristic time of 
diffusion, $\tau_{ad}<<\tau_{D}$, we may assume that the adsorbed phase is
practically in equilibrium. Under this condition, one has from (\ref{kinetic-langmuir})
\begin{equation}\label{langmuirgen}
c_1=\alpha\frac{c_2}{Z(c_2)} ,
\end{equation}
where $\alpha=k_d/k_a$. For the case of a non-equilibrium stationary state 
in which $d c_2/dt=J_{ads}$, with $J_{ads}$ a constant current of particles, 
one obtains
\begin{equation}\label{langmuirgen}
c_1=\alpha\frac{1}{Z(c_2)}\left({c_2} + \frac{J_{ads}}{k_a}\right).
\end{equation}

\begin{figure}[tbp] 
\begin{center}
\includegraphics[scale=0.40]{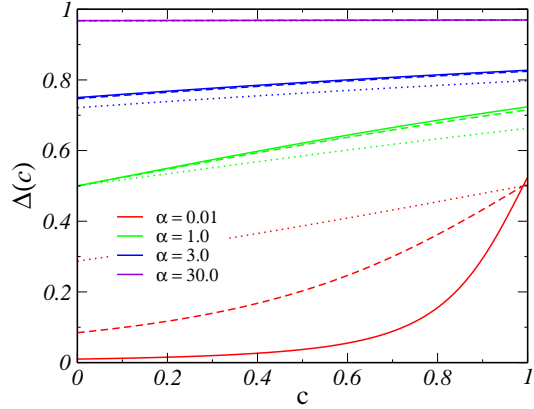} 
\caption{The factor $\Delta(c)$ from (\ref{deltaclangmuir}) in terms of concentration
for different values of the parameter $\alpha$. Solid lines correspond to equilibrium 
adsorption (\ref{langmuir}) whereas dashed and dotted lines correspond to the non-equilibrium 
stationary case (\ref{deltaclangmuirstationary}) for $\tilde{J}=0.1cm^{-3}$ and $\tilde{J}=1.1cm^{-3}$, 
respectively. The values of $\alpha$ used were (from red to purple) 
$\alpha=k_d/k_a=0.01,\,1.0,\,3.0 \text{ and}\,30.0$ and $\xi=1cm^{3}$. }
\label{fig2}
\end{center}
\end{figure}

\subparagraph*{Low coverage:} 
When the concentration of the adsorbed particles is small we may assume $Z_2 \simeq 1- \xi c_2$, with $\xi=1/c_2^{max}$ the
inverse value of the total concentration of adsorbing sites $c_2^{max}$ in the system\cite{rubi01,roque}. Then, 
Eq. (\ref{kinetic-langmuir}) yields the Langmuir isotherm \cite{rubi01} 
\begin{equation}
c_2=\phi(c_1)=\frac{c_1}{\alpha+ \xi c_1}. \label{langmuir} 
\end{equation}
Using Eq. (\ref{langmuir}) in the condition $c=c_1+\phi(c_1)$, one obtains
\begin{equation} \label{c1delangmuir}
 c_1(c)=\frac{1}{2 \xi}\left(\overline{c} -\alpha+\sqrt{(\overline{c} -\alpha-1)^2+4\overline{c} \alpha}-1\right), 
\end{equation}
where $\overline{c}\equiv c \xi$ is the total concentration scaled by the total concentration of adsorbing sites. From Eq. (\ref{c1delangmuir}), 
the correction factor $\Delta(c)$ becomes
\begin{equation} \label{deltaclangmuir}
 \Delta(c)=\frac{1}{2}\left(1+\frac{\overline{c}-\alpha-1+2\alpha}{\sqrt{(\overline{c}-\alpha-1)^2+4\overline{c}\alpha}}\right). 
\end{equation}
In Fig \ref{fig2}, the correction factor $\Delta(c)$ is represented as a function of the total 
concentration $c$ for different values of $\alpha$ in the equilibrium case. From now on,
we will assume for simplicity that $\xi=1cm^{3}$. Values of $\xi$ different from the unity only rescale the results, the
qualitative behavior of the functions remains the same. When adsorption 
is dominant, as occurs for $\alpha \sim 0.01$, the effective diffusion coefficient becomes 
negligible for concentrations $c < 0.4cm^{-3}$. Increasing the importance of desorption, 
$\alpha \sim 1.0$, makes $\Delta(c)$ more significant taking values of about $50\%$ 
to $70\%$ of $D_1(c)$. As expected, this tendency becomes
even more manifest when desorption is faster than adsorption, $\alpha >> 1$.

In similar form, for the stationary case we have
\begin{equation}
c_2=\phi_s(c_1)=\frac{c_1}{\alpha+c_1}-\frac{J_{ads}}{k_a}\frac{1}{\alpha+c_1}. \label{langmuirNeq}
\end{equation}
and therefore the correction factor $\Delta_{s}(c)$ is in this case
\begin{equation} \label{deltaclangmuirstationary}
\Delta_{s}(c)=\frac{1}{2}\left(1+\frac{c-\alpha-1+2\alpha}{\sqrt{(c-\alpha-1)^2+4(\tilde{J}+c\alpha)}}\right). 
\end{equation}
here we have introduced the parameter $\tilde{J}=J_{ads}/k_a$, having dimensions of number concentration.

Dashed and dotted lines in Fig \ref{fig2} show the stationary non-equilibrium values of
$\Delta(c)$ as a function of the total concentration $c$ for the same values of $\alpha$ in the 
equilibrium case and for $\tilde{J}=0.1cm^{-3}$ (dashed) and $\tilde{J}=1.1cm^{-3}$ (dotted).  The 
stationary adsorption process is still dominant for $\alpha \sim 0.01$, but in 
this case the effective diffusion coefficient takes values considerably larger than 
those of the equilibrium case, $10\%$ for $J_{ads}=0.1cm^{-3}$ and $30\%$ for $J_{ads}=1.1cm^{-3}$. As in 
the equilibrium case, this effect decreases when increasing $\alpha$. This behavior of the correction factor $\Delta(c)$ 
as a function of particle concentration in the bulk is similar to the one observed for the effective diffusion coefficient 
in experiments and simulations \cite{rutven,suffriti1,suffriti2}. The relation between both quantities can be established by means of
Eq. (\ref{imm_flux2}). Unfortunately, a quantitative comparison is not possible due to the fact that reported results in simulations
are given in terms of the so-called loading parameter measuring the number of particles in the pore.

\subparagraph*{High coverage:} When the concentration of adsorbed particles is higher,
one has to take into account nonlinear contributions to $Z(c_2)$ through the virial expansion
\begin{equation}\label{Z2interaccion}
Z(c_2)\simeq 1- \xi c_2+bc_2^2 + O(c_2^3),
\end{equation}
where $b$ is a constant having dimensions of square volume and may take positive or negative values 
depending on the nature of the interactions between adsorbed-adsorbed particles and may also depend on temperature. 
It is convenient to mention that the expansion (\ref{Z2interaccion}) is consistent with a second order virial expansion in which interactions 
(excluded volume and direct ones) between adsorbed particles are taken into account. This is the case since the factor $Z(c_2)$ plays the role of a fugacity. 
More important is that Eq. (\ref{Z2interaccion}) allows us to go beyond the classical Langmuir model. 
Using this relation, we obtain the following adsorption isotherm
\begin{figure}[tbp]
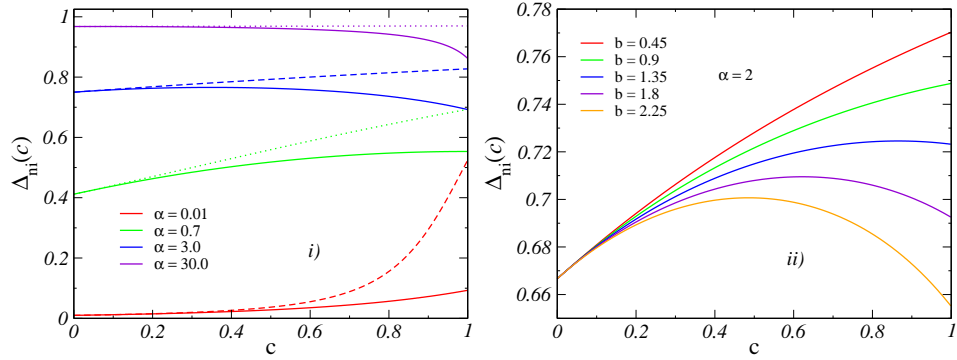
 
\begin{center}
\includegraphics[scale=0.35]{Grzywna-Fig2a.eps} 
\includegraphics[scale=0.35]{Grzywna-Fig2b.eps} 
\caption{The nonideal correction factor $\Delta_{ni}(c)$ in terms of the total concentration $c$,
for $b=(1+\alpha)^2/4$ and different values of $\alpha=k_d/k_a=0.01,\,1.0,\,3.0 \text{ and }\,30.0$ 
(from red to purple)  and $\xi=1cm^{3}$. \textit{i}) The solid lines represent the behavior of  $\Delta_{ni}(c)$ for 
equilibrium adsorption whereas dashed lines correspond to the ideal case. Interactions between 
the adsorbed particles introduce a nonmonotonic behavior of the effective diffusion 
coefficient for $ \alpha \geqslant 1$, which depends on the value of $b$.
\textit{ii}) The nonideal correction factor for $\alpha = 2$,  $\xi=1cm^{3}$ and 
$4b/(1+\alpha)^2=0.2,\,0.4,\,0.6,\,0.8 \text{ and }1.0 cm^{6}$ (from red to orange).}
\label{fig3}
\end{center}
\end{figure}

\begin{equation}\label{langmuirNoIdeal}
c_2=\phi_{ni}(c_1)=\frac{\alpha+c_1 \xi}{2bc_1}\left[ 1\pm \sqrt{1-{4bc_1^2}/{(\alpha+c_1  \xi)^2}}\right], 
\end{equation}
where the subindex $ni$ indicates the non ideal case under consideration. Note that 
Eq. (\ref{langmuirNoIdeal}) is valid for $b>0$. Substitution of this equation into the mass 
balance leads to the relation 
\begin{equation} \label{c1NoIdeal}
 c=c_1+\frac{\alpha+c_1 \xi}{2bc_1}\left[ 1\pm \sqrt{1-{4bc_1^2}/{(\alpha+c_1  \xi)^2}}\right]. 
\end{equation}
This equation is a third degree polynomial equation for $c_1(c)$ that has a 
real solution from which the correction factor $\Delta_{ni}(c)$ can be calculated. The 
resulting relations can be used to estimate the corrections 
due to the non-ideal nature of the adsorbed phase. The condition $b \leq (1+\alpha)/2$,
obtained from Eqs. (\ref{langmuirNoIdeal}) and (\ref{c1NoIdeal}), guarantees the existence of 
real values for the concentration $c$.

Figure \ref{fig3}\textit{i}) shows the nonideal correction factor  $\Delta_{ni}(c)$ in terms of the total 
concentration $c$ by assuming $\xi=1 cm^{3}$. For $\alpha \lesssim 2$, interactions reduce the magnitude of the effective 
diffusion coefficient although it increases monotically with $c$. However, for 
$\alpha \geqslant 1$, interactions between particles of the adsorbed phase may 
cause a non-monotonic behavior of the effective diffusion coefficient for certain values 
of $b$, making it to decrease for larger values of $c$ as it is 
shown in detail in Figure \ref{fig3}\textit{ii}) for which $\alpha =2$. This behavior implies 
that an optimal value for the concentration exists for which the effective diffusion coefficient 
reaches its maximum value. For $\alpha > 10$, interactions monotonically reduce the 
effective diffusion coefficient as $c$ increases. As in the previous case, the non-monotonic behavior 
manifested by the correction factor  agrees well with the non-monotonic behavior of the effective diffusion 
coefficient observed in simulations and experiments\cite{rutven,suffriti1,suffriti2,suffriti3}. 

\section*{3. The effect of surface diffusion on the adsorption-diffusion interplay}
The case in which the adsorbed particles may diffuse on the surface of the 
sorbent can be analyzed by adding a diffusion term to Eq. (\ref{kinetic-langmuir}), \cite{rubi01}. 
The evolution equation for $c_2$ takes the form
\begin{equation}\label{kinetic-langmuir2}
\frac{\partial}{\partial t} c_2 = k_a c_1 Z(c_2) - k_dc_2  
+\frac{\partial}{\partial x}\left[ D_2(c_2) \frac{\partial}{\partial x}  c_2\right],
\end{equation}
where $D_2(c_2)$ is the diffusion coefficient at the surface.

An estimation of the contribution of surface diffusion to the
adsorption-diffusion interplay can be performed by assuming that 
$D_2 \sim L^2 /\tau_D$ where $\tau_D$ is the characteristic diffusion time  
and $L$ a characteristic length of the sorbent surface.
In first approximation, the diffusion term in (\ref{kinetic-langmuir2}) can be
expressed as
\begin{equation}\label{langmuir2-dif}
\frac{\partial}{\partial x}\left[ D_2(c_2) \frac{\partial}{\partial x}  c_2\right] 
\sim \frac{c_2}{\tau_D}.
\end{equation}
This type of approximation is frequently and succesfully used in transport processes problems 
to examine and estimate, for instance, viscous or thermal effects near to a surface.

At low coverage, when the adsorbed phase is dilute, one can make the appproximation 
$Z(c_2) \simeq 1- \xi c_2$ . Eq. (\ref{kinetic-langmuir2}) reduces to
\begin{equation}\label{kinetic-langmuir3}
\frac{\partial}{\partial t} c_2 = k_a c_1 (1-\xi c_2) - k_d^{eff}  c_2, 
\end{equation}
where $ k_d^{eff} =k_d\left(1 -\tau_D^{-1}k_d^{-1}\right)$. This equation leads to the
adsorption isotherm 
\begin{equation}\label{langmuirIsotherm3}
c_2=\phi(c_1)=\frac{c_1}{\tilde{\alpha}+\xi c_1 } ,
\end{equation}
where now $\tilde{\alpha}=k_d^{eff} /k_a<\alpha$. Therefore, in the dilute case the influence 
of surface diffusion on adsorption is to reduce the desorption rate. As a 
consequence, the values of the effective diffusion coefficient of the system $D_{eff}(c)$ become 
shifted to lower concentrations.

In the nondilute case, both $Z(c_2)$ and $f_2$ must include
non-linear contributions. The expression of $Z(c_2)$ is given in Eq. (\ref{Z2interaccion}). 
According to Eq. (\ref{termodDf}), the activity coefficient can be written in terms of
the non-ideal contribution of the chemical potential: $f_2=exp(\Delta\mu_{ni}/RT)$. A virial expansion
of $\Delta\mu_{ni}$ at the lower order in concentration leads to the relation $f_2\simeq exp(\beta c_2)$, 
where $\beta = \tilde{B}_2/RT$ has dimensions of volume and $\tilde{B}_2(T)$ is the second virial coefficient which must 
take into account excluded volume effects and direct interactions. 
\begin{figure}[tbp]
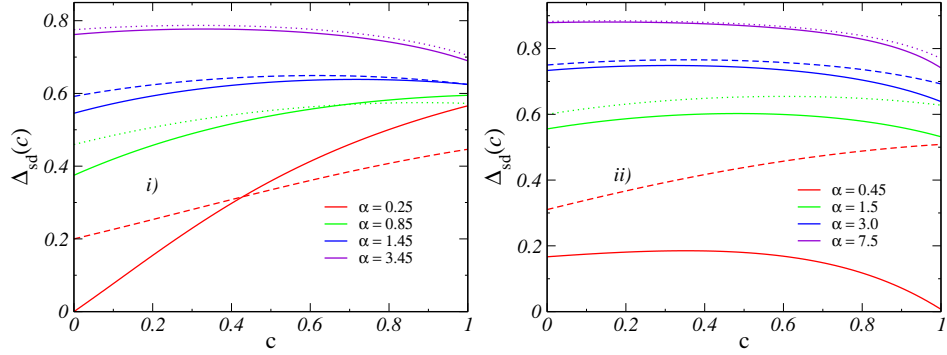
 
\begin{center}
\includegraphics[scale=0.35]{Grzywna-Fig3a.eps} 
\includegraphics[scale=0.35]{Grzywna-Fig3b.eps} 
\caption{The nonideal correction factor for the case of surface diffusion $\Delta_{sd}(c)$ in terms of 
the total concentration $c$ for $b=(1+\alpha)^2/4$, $\tilde{J}=0$, $\gamma=\tau_d^{-1}k_a^{-1}=0.25$ and different 
values of $\alpha=k_d/k_a$ and $\beta$. In \ref{fig4}\textit{i}) the solid lines represent the behavior of  
$\Delta_{sd}(c)$ for equilibrium adsorption whereas dashed lines correspond to $\Delta_{ni}(c)$ 
for $\beta=-1.5 cm^{3}$ (low temperatures) and $\alpha=0.25,\,0.85,\,1.45,\,3.45 \text{ and }\,13.45$ 
(from red to purple). The nonmonotonic behavior of the effective diffusion coefficient is more 
pronounced for $\alpha <1$.  \textit{ii}) shows $\Delta_{sd}$ 
for $\beta=0.5cm^{-3}$ (high temperatures) and $\alpha=0.45,\,1.5,\,3.0,\,7.5 \text{ and }\,15.0$ 
(from red to purple) and $\xi=1cm^{3}$. For the present model, high values of $\beta >0$ may induce a discontinuity of 
the correction factor.}
\label{fig4}
\end{center}
\end{figure}

Using the previous expressions for $Z(c_2)$ and $f_2(c_2)$, one obtains the generalized
Langmuir equation
\begin{equation}\label{kinetic-langmuir4}
 \frac{\partial}{\partial t} c_2 = k_a c_1 \left(1-\xi c_2+bc_2^2\right) - k_d^{eff} c_2 ,
\end{equation}
where now the effective desorption rate is 
$k_d^{eff}=k_d\left[1 - \tau_d^{-1}k_d^{-1}\left(1+\beta c_2\right)\right]$. 
The coefficient $\beta$ can be positive or negative depending on the temperature. 
In the non-equilibrium stationary case one obtains from
Eq. (\ref{kinetic-langmuir4}) 
\begin{equation} \label{c1NoIdealdiff}
c=c_1+\frac{\alpha-\gamma+ \xi c_1}{\beta \gamma+2bc_1}
\left[ 1\pm \sqrt{1-{4(c_1-J)(bc_1+\beta \gamma)}/{(\alpha-\gamma+\xi c_1)^2}}\right]. 
\end{equation}
where $\gamma=\tau_d^{-1}k_a^{-1}$. 

In Figure \ref{fig4}, we show the nonideal correction factor $\Delta_{sd}(c)$ for $\xi=1$ in the case of 
equilibrium adsorption with surface diffusion in terms of the total concentration 
$c$ for low, Fig. \ref{fig4}\textit{i}), and high temperatures, Fig. \ref{fig4}\textit{ii}). 
As in Fig. 2, interactions induce 
a non-monotonic behavior but in the present case, for $\alpha < 1$ and low temperatures ($\beta<0$), 
they lead to an enhancement of diffusion for large concentrations due to the attractive nature of 
direct interactions of the particles in channel. At low concentrations, diffusion is significatively 
lower than in the case of Fig. 2. When the concentration is higher and the temperature
is such that $\beta>0$, excluded volume effects cause suppression of diffusion,
as observed in the figure.

\section*{4. Conclusions}
In this article, we have analyzed the combined effect of diffusion and adsorption
in porous materials by adopting an effective medium description based on non-equilibrium 
thermodynamics \cite{rubi01}.
 
We have treated the case of very narrow pores, present in some porous materials, microfluidic devices and in biological channels, in which particles move essentially along the direction of the pores and may occasionally become attached to their surface. The fact that the pores are narrow, makes it possible to adopt a quasi-onedimensional description in terms of an average concentration along the channel and a kinetic mechanism governing the attachment-detachment of the particles to the surface of the channel. The model proposed considers the combined effect of these two mechanisms and analyzes particle transport by means of effective diffusion in which the diffusion coefficient depends on the adsorption kinetics described in turns by a generalized Langmuir equation.
 
The analysis of the effective diffusion coefficient reveals the existence of a multiplicity of different regimes
controlled by different adsorption-desorption kinetics and different coverages of particles at the 
surface. We have quantified the change of the diffusion coefficient and found a non-monotonic behavior 
of this quantity due to excluded volume interactions. These results are in qualitative agreement
with previous ones obtained from numerical models.
 
The model proposed could be extended to the case of different shapes of the pores by considering entropic 
barriers emerging from the variation of the cross section of the pores. This generalization can be 
performed following the approach presented in Refs. \cite{entropBarrier1,entropBarrier2}. The 
resulting model may represent a simple and useful tool to analyze particle transport in nanoporous 
materials and biological channels.

\section*{Acknowledgements}
This work was supported by the DGiCYT of Spanish Government under Grant No. FIS2008-04386 
and by UNAM DGAPA Grant No. IN100112-2. ISH appreciates the hospitality of Prof. J. M. Rubi and 
the Department of F\'isica Fonamental of the University of Barcelona where this work was 
done during a sabbatical leave. J. M. Rubi acknowledges financial support from 
Generalitat de Catalunya under program Icrea Academia. Z.J.Grzywna acknowledges support from MNiSzW under the project NN508409137.

\end{document}